\documentclass[%
byrevtex,
prd,
showpacs,
showkeys,
nofootinbib
]{revtex4}
\usepackage[pdftex]{graphicx}
\usepackage{amsmath,amssymb,amsxtra}
\usepackage{bm}
\usepackage{float,color} 
\begin{document} 
\title{Analog rotating black holes in a magnetohydrodynamic inflow}
\author{Sousuke Noda}
\email{noda@gravity.phys.nagoya-u.ac.jp}
\author{Yasusada Nambu}%
\email{nambu@gravity.phys.nagoya-u.ac.jp}
 \affiliation{%
  Department of Physics, Graduate School of Science, Nagoya University,
  Chikusa, Nagoya 464-8602, Japan}
\author{Masaaki Takahashi}
\email{takahasi@phyas.aichi-edu.ac.jp}
\affiliation{Department of Physics and Astronomy, Aichi University of
  Education, Kariya, Aichi 448-8542, Japan}
 
\date{March 1, 2017}
\begin{abstract}
  We present a model of the analog geometry in a magnetohydrodynamic
  (MHD) flow.  For the MHD flow with magnetic pressure-dominated and 
  gas pressure-dominated conditions, we obtain the magnetoacoustic metric for
  the fast MHD mode.  For the slow MHD mode, on the other hand,
  the wave is governed by the advective-type equation without an
  isotropic dispersion term. Thus, the ``distance'' perpendicular
  to the wave propagation is not defined and the magnetoacoustic
  metric cannot be introduced. To investigate the properties of the
  magnetoacoustic geometry for the fast mode, we prepare a 
  two-dimensional axisymmetric inflow and examine the behavior of 
  magnetoacoustic rays which is a counterpart of the MHD waves in the
  eikonal limit. We find that the magnetoacoustic geometry is
  classified into three types depending on two parameters
  characterizing the background flow:~analog spacetimes of rotating
  black holes, ultra spinning stars with ergoregions, and spinning
  stars without ergoregions. We address the effects of the magnetic
  pressure on the effective geometries.
\end{abstract}

\maketitle


\section{introduction}
Black holes predicted by Einstein's general relativity are
characterized by the event horizon from which even light rays cannot
escape.  A rotating black hole has the ergoregion where
light rays cannot propagate to the counterrotating direction. When we
consider the wave scattering, rotating black holes can amplify waves, 
and this phenomenon is called superradiance \cite{Zeldovich1971,
  Zeldovich1972, Brito2015a}.  From the analysis of the wave scattering, it turns
out that superradiance with some boundary conditions can evoke
instabilities of waves.  For example, considering the scattering of
massive fields in the Kerr spacetime or massless fields in the
Kerr-AdS spacetime, the wave is confined around the black hole due to
a wall of effective potential for waves, and the superradiant
scattering occurs repeatedly. Then the amplitude of the wave finally
grows exponentially in time.  This kind of instability is called the
superradiant instability \cite{Teukolsky1972}.  Even if
 there is no potential wall for confinement, the wave also suffers
from the instability if there exists a reflective inner boundary such
as the surface of stars (ergoregion instability)~\cite{Friedman1978,
  Comins1978, Vilenkin1978, Cardoso2008, Glampedakis2013, 
  Cardoso2008a, Pani2010}.

To examine the properties of superradiance associated with rotating
black holes, we can use analog models of black holes.  
Analogs based on fluid models are called acoustic black holes.  If a
flow of the fluid has a sonic surface, the acoustic wave
cannot pass through the sonic surface from the downstream to the
upstream.  Namely, the sonic surface behaves as the black hole horizon
for the acoustic wave (acoustic horizon).  In such a
  situation, the perturbation of the velocity potential of the
  fluid obeys the Klein-Gordon equation in a ``curved" spacetime of
which geometry is defined by the acoustic metric~\cite{Unruh1981}.
That is why this kind of analog model is called an acoustic black hole
(see also Ref.~\cite{Barcelo2011}).  Although the motivation of the
original work by Unruh \cite{Unruh1981} was to investigate the Hawking
radiation from an acoustic black hole in laboratories, some papers later 
examined the feature of superradiance for the
acoustic wave (acoustic superradiance) as well.  To realize the
rotating acoustic black hole in laboratories, several experimental
setups or models have been proposed \cite{Visser1998, Schutzhold2002,
  Basak2003, Visser2005, Richartz2009, Richartz2015, Dolan2013, 
  Oliveira2014, Oliveira2015, Cardoso2016, Berti2004, Dolan2012}.  The draining bathtub model~\cite{Visser1998} is one
example of such flows, and its acoustic metric has a structure
  similar to a rotating black hole.  In this model, the flow is 
 two dimensional, and the acoustic
horizon is a closed circle where the radial velocity of the background
flow exceeds the sound velocity in the fluid. The analog
structure of the ergoregion also exists.

There are several works that consider the acoustic black holes in
astrophysical situations. The original idea of the acoustic
geometry in relativistic fluids was proposed by 
Moncrief \cite{Moncrief1980} to examine the stability of the accretion 
flow onto a Schwarzschild black hole. 
In Moncrief's analysis, the acoustic wave in the accretion
flow was shown to satisfy the Klein-Gordon-type wave equation.  Some
papers also apply the idea of the acoustic black holes to astrophysical
phenomena \cite{Bilic1999, Abraham2006, DasBilic2006, Das2008,
  Tarafdar2013, Pu2012, Ananda2014, Ananda2015}. 
  In particular, Abraham {\textit{et al.}}~\cite{Abraham2006} study 
  the axially symmetric accretion flow onto a Kerr black hole and 
  discuss a situation similar to the draining bathtub model with acoustic superradiance.
When we apply the concept of acoustic black holes to some 
astrophysical problems, we should consider the effects of the magnetic field. 
In fact, due to the angular-momentum transportation caused by the
magnetic viscosity, matter in the accretion disk can fall efficiently
onto the central compact object \cite{Balbus1991, Balbus2002,
  Balbus2003, Yokosawa2005}.  Hence, when we discuss acoustic black holes 
  in magnetized fluids such as in accretion disks, we need to take into
account the magnetic field and generalize the acoustic black hole
to the magnetohydrodynamic (MHD) case.

In this paper, we discuss the acoustic geometry in a MHD flow
(magnetoacoustic geometry).  As the eikonal equation of the MHD waves
is the fourth-order differential equation for the phase of the
velocity perturbation, it is not possible to define the quadratic
magnetoacoustic metrics, and the MHD wave equation cannot be written in
the form of the Klein-Gordon equation in curved spacetimes.  To
circumvent this problem, we focus on the magnetic pressure-dominated and 
  the gas pressure-dominated cases of magnetized fluids.  In such restricted
cases, it is possible to expand the eikonal equation to define the
quadratic magnetoacoustic metric.  As a model of background
 flows, we prepare a stationary and axisymmetric two-dimensional
ideal MHD inflow with a sink at the center.
  In our model, the magnetoacoustic property is characterized by two 
  parameters relating to the sound velocity in the fluid and the angular 
  velocity of the magnetic field line at the radial Alfv\'en point
  where the radial velocity of the fluid coincides with the radial component 
  of the Alfv\'en velocity.
  Using these parameters, the magnetoacoustic geometry is classified into three types 
  according to the values of these two parameters.

  This paper is organized as follows.  In Sec.~II, we review one of
  the draining bathtub-type models and the superradiant scattering of
  the acoustic waves.  In Sec.~III, we first define the
  magnetoacoustic metric by considering the magnetic pressure-dominated
  case and the gas pressure-dominated case.  Then we introduce the stationary
  and axisymmetric background MHD flow.  In Sec.~IV, we present
  possible magnetoacoustic geometries for our MHD flow.  Section V is
  devoted to a summary of our paper.  We use units $c=G=\hbar=1$
  throughout this paper.


\section{Acoustic black hole for perfect fluid}
\subsection{Draining bathtub model}

We review an analog rotating black hole with 
two-dimensional flows, which is essentially the same as the draining bathtub
model~\cite{Visser1998}.  We consider an irrotational perfect fluid in
a flat spacetime.  The basic equations are
\begin{align}
&\frac{\partial \rho}{\partial t}+\nabla\cdot(\rho\, \bm{v
})=0,
\label{eq:continu}\\
&\frac{\partial \bm{v}}{\partial t}+(\bm{v} \cdot
  \nabla)\bm{v}+\frac{\nabla p}{\rho}+\bm{F}_\text{ex}=0,
\label{eq:eulereq}\\
&\bm{v}=\nabla \Phi,
\label{eq:velo}
\end{align}
where the fluid velocity $\bm{v}$ is represented by the velocity
potential $\Phi$ and $\bm{F}_{\text{ex}}$ is an external force to
control the flow.  We assume the pressure $p$ and the density $\rho$
of the fluid obey the polytropic equation of state
$p\propto \rho^\Gamma$.
 We consider an
axisymmetric stationary inflow and introduce the polar coordinates
$(R, \phi)$. Then the velocity and the external force are
\begin{equation}
  \bm{v}=v^R(R)\,\bm{e}_{\hat R}+v^\phi(R)\,\bm{e}_{\hat\phi},\quad
  \bm{F}_\text{ex}=F_\text{ex}(R)\,\bm{e}_{\hat R},
\end{equation}
where $\bm{e}_{\hat{R}}$ and $\bm{e}_{\hat{\phi}}$ are the orthonormal
basis vectors. As the flow is two dimensional, it does not have a 
$z$ component ($v^{z}=0$, $F_\text{ex}^z=0$). 
From the continuity equation \eqref{eq:continu}, we have
  \begin{equation}
      R\,\rho\, v^R=\text{const}, \label{eq:contSol}
  \end{equation}
and from the azimuthal component of Eq.~\eqref{eq:eulereq},
\begin{equation}
  R\,v^\phi=\text{const}\equiv L,
\end{equation}
where $L$ is the angular momentum of the flow. The radial component
of Eq.~\eqref{eq:eulereq} is
\begin{equation}
  v^R\frac{dv^R}{dR}-\frac{v^\phi}{R}+\frac{1}{\rho}\frac{dp}{d\rho}+F_\text{ex}
=0.
\label{eq:ber}
\end{equation}
This equation yields the relation between $v^{R}$ and $F_\text{ex}$
because the pressure is related to the density as
$p\propto \rho^{\Gamma}$ and the density is described by $v^{R}$ and
$R$ through the relation \eqref{eq:contSol}.  For a given radial
velocity profile $v^R(R)$, this relation defines an
appropriate form of $F_\text{ex}(R)$.  We choose the radial velocity
as
\footnote{In some papers on the draining bathtub model, for example
  \cite{Visser1998}, the radial velocity is chosen as $v^{R} \propto
  R^{-1}$ to make $\rho=$const. 
The reason why we choose the nonstandard radial velocity is that we will
compare it with our MHD model discussed in Sec.III which has the
radial velocity $\propto R^{-1/2}$.}
\begin{equation}
  \label{eq:flow0}
  v^R\propto R^{-1/2}.
\end{equation}
The sound velocity is
\begin{equation}
  c_s \equiv \sqrt{\frac{\partial p}{\partial \rho}}\, \propto \Gamma^{1/2} R^{-(\Gamma-1)/4}.
  \label{eq:sonic}
\end{equation}
To define an acoustic black hole, the background flow has to be a
transonic inflow, and the downstream of the sonic point needs to be 
supersonic.
Since the sonic point $R=R_s$ is given by
\begin{equation}
  c_s=|v^R|,
  \label{eq:ac_hor}
\end{equation}
we consider the polytropic index within the interval $0<\Gamma <3$ 
for the radial velocity \eqref{eq:flow0}.
By normalizing all quantities at the sonic point 
$R_s$, we obtain $\bm{v}$
and $c_s$ as functions of $R$ as
\begin{equation}
  \label{eq:bathtub}
  v^R=-c_{s}(R_s)\left(\frac{R}{R_s}\right)^{-1/2},\quad
  v^\phi=\frac{L}{R},\quad
  c_s=c_{s}(R_s)\left(\frac{R}{R_s}\right)^{-(\Gamma-1)/4}.
\end{equation}
Figure 1 shows the structure of this flow.
\begin{figure}[H]
\centering
\includegraphics[width=0.6\linewidth]{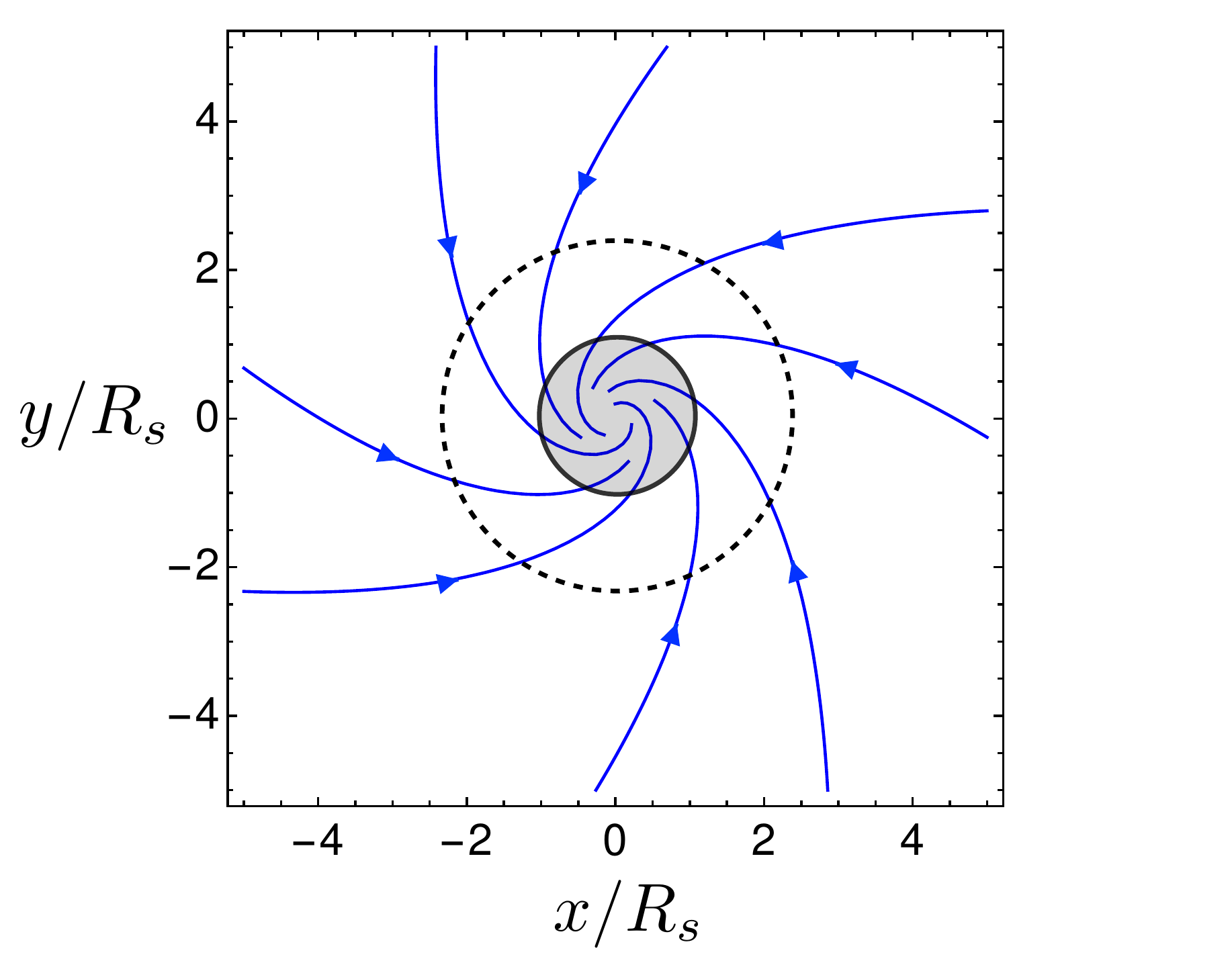}
\caption{Draining bathtub flow in the $xy$ plane with $\Gamma=4/3$,
     $c_s(R_s)=1,~{L}=1.2R_s$, where $R_s$ is
  the radius of the acoustic horizon defined by
  Eq.~\eqref{eq:ac_hor}.  The solid curves are streamlines of the
  flow.  The grey region represents the inside of the acoustic horizon, and
  the dotted circle is the ergosurface for the acoustic waves.}
\label{fig:bathtubflow}
\end{figure}

Now, we add small perturbations $\delta\bm{v}, \delta\Phi, \delta\rho$
to the background flow.  We assume that the perturbations are
independent of the $z$ coordinate.  Combining the first-order
perturbed equations of \eqref{eq:continu}-\eqref{eq:velo}, 
the perturbation of the velocity potential
$\delta \Phi$ satisfies the following wave equation:
\begin{equation}
\frac{\partial }{\partial t}\left[c_{s}^{-2}\rho
\left(\frac{\partial \delta \Phi}{\partial t}+\bm{v}\cdot \nabla 
\delta \Phi \right)\right]+ \nabla
\cdot\left\{ \left[c_{s}^{-2}\rho\left(\frac{\partial \delta
              \Phi}{\partial t}+\bm{v}\cdot \nabla \delta \Phi 
\right)\right] \bm{v} -\rho\nabla \delta \Phi \right\}=0\ \ ,\ \ 
\frac{\partial \delta \Phi}{\partial z}=0.
\label{eq:dens_2}
\end{equation}
Equation \eqref{eq:dens_2} can be written as
\begin{equation}
\frac{\partial}{\partial x^{\mu}}
\left(f^{\mu\nu}\frac{\partial \delta \Phi} {\partial x^{\nu}}\right)=0,\quad\mu,\nu=0,1,2,
\label{eq:f}
\end{equation}
with the matrix $f^{\mu\nu}$ defined by
\begin{equation} 
f^{\mu\nu}(t, \bm{x})=
\frac{\rho}{c_{s}^2}
\begin{pmatrix}
 -1 & -v^i \\ -v^i & c_s^2\,\delta^{ij}-v^i\,v^j
\end{pmatrix},\quad i,j=1,2.
\label{eq:f_comp}
\end{equation}
We introduce a matrix $s^{\mu\nu}$ as
\begin{equation}
\sqrt{|s|}\,s^{\mu\nu}=f^{\mu\nu},\quad s=\mathrm{det}\,s_{\mu\nu}.
\label{eq:g}
\end{equation}
The matrix $s^{\mu\nu}$ and its inverse are given by
\begin{equation}
s^{\mu\nu}(t, \bm{x})=\frac{1}{\rho\, c_{s}}
\begin{pmatrix}
 - 1 & -v^{j} \\ -v^{i}  & c_{s}^2\,\delta^{i
   j}-v^{i}\,v^{j}
\end{pmatrix}
,\quad 
 s_{\mu\nu}(t, \bm{x})=\frac{\rho}{c_{s}}
\begin{pmatrix}
 - (c_{s}^2-v^2) & -v^{j} \\ -v^{i}  &  \delta^{ij}
\end{pmatrix}.
\label{eq:acmetric}
\end{equation}
Then, Eq.~\eqref{eq:dens_2} becomes
\begin{equation}
\frac{1}{\sqrt{|s|}}\frac{\partial}{\partial x^{\mu}}
\left(\sqrt{|s|}\,s^{\mu\nu}\frac{\partial
 \delta \Phi}{\partial x^{\nu}}\right)=\square_s \delta \Phi=0,
\label{eq:s-eq}
\end{equation}
where the d'Alembertian $\square_s$ is defined with the metric
$s_{\mu\nu}$. Equation \eqref{eq:s-eq} is the Klein-Gordon equation for a
scalar field $\delta\Phi$ in the curved spacetime with the ``acoustic
metric'' $s_{\mu\nu}$.  This metric represents the effective
geometry for the acoustic waves. For transonic background flows, it
can be shown that the geometry represented by $s_{\mu\nu}$ has a
similar structure to black hole spacetimes.  The acoustic interval
(acoustic line element) is defined as
\begin{equation}
ds^2 \equiv s_{\mu\nu}dx^{\mu}dx^{\nu}
=\frac{\rho}{c_{s}}\Bigl[-(c_{s}^2-\bm{v}^2)dt^2
 -2\bm{v} \cdot d \bm{x}\,dt +d\bm{x}^2\Bigr]\ \ ,\ \ d\bm{x}=(dx, dy).
\label{eq:acmetric1}
\end{equation}
In the polar coordinates $(R, \phi)$, 
the acoustic interval~\eqref{eq:acmetric1} becomes
\begin{equation}
ds^2=\frac{\rho}{c_s}\left\{-\left[c_s^2-\left((v^{R})^2
+(v^{\phi})^2\right)\right]dt^2-2v^{R}\,dt\,dR-2v^{\phi}\,R\,dt\,d\phi
+dR^2+R^2d\phi^2 \right\}.
\label{eq:line_polar}
\end{equation}
 To examine the characteristics of this geometry, 
 we rewrite the metric~\eqref{eq:line_polar} by applying the
  following coordinate transformation:
\begin{equation}
dt\rightarrow dt+\frac{v^{R}}{c_s^2-(v^{R})^2}\,dR,\quad
d\phi \rightarrow d\phi+\frac{v^{R}\,v^{\phi}}{c_s^{2}-(v^{R})^2}\,
\frac{dR}{R}.
\label{eq:new_coordinate}
\end{equation}
Then we can write the acoustic interval in the following form:
\begin{equation}
ds^2=\frac{\rho}{c_{s}}\left\{-\left[c_{s}^2-\left((v^{R})^2
+(v^{\phi})^2\right)\right]dt^2-2v^{\phi}\,R\,d{\phi}\,d{t}
+\frac{dR^2}{1-(v^R/c_s)^2}
+R^2\,d{\phi}^2\right\},
\label{eq:bathtub_line_element}
\end{equation}
where we used $(t,R,\phi)$ as the new
coordinates after the coordinate transformation
\eqref{eq:new_coordinate}.  From the similarity between the acoustic
metric \eqref{eq:bathtub_line_element} and the metric of rotating
black holes, we can introduce the acoustic horizon and ergosurface.
The $RR$ component and the $tt$ component of the acoustic metric
provide the following conditions:
\begin{align}
&\left |v^{R}\right |=c_{s} &  & \text{(acoustic~horizon)},\\ 
&\sqrt{(v^{R})^2+(v^{\phi})^2}=c_{s}&  & \text{(acoustic~ergosurface)}.
\label{eq:ac_hor_ergo}
\end{align}
The acoustic horizon is a circle where the radial velocity $v^R$
coincides with the sound velocity and the acoustic wave cannot
propagate outward from the inside of the acoustic horizon.  The
acoustic ergoregion is defined by
$\sqrt{(v^R)^2+(v^\phi)^2}\ge c_s$, and the
wave propagation against the flow is not possible.  For the background
flow~\eqref{eq:flow0}, the radius of the acoustic horizon
$R_\text{H}$ is
$R_s$ (the sonic point, see Fig.~\ref{fig:bathtubflow}).
The radius of the ergosurface $R_\text{E}$ is determined as
the solution of the following equation:
\begin{equation}
 \left(\frac{R}{R_s}\right)^{(5-\Gamma)/2}-\left(\frac{R}{R_s}\right)-\left(
   \frac{L}{c_{s}(R_s)R_s}\right)^2=0.
\end{equation}


\subsection{Acoustic superradiance in the draining bathtub flow}
We consider the solution of the Klein-Gordon equation~\eqref{eq:s-eq} with the
draining bathtub flow~\eqref{eq:bathtub}.  The wave function
$\delta \Phi$ can be separated as
\begin{equation}
\delta\Phi
=\frac{\psi(R)}{ R^{(7-\Gamma)/16}}\,e^{i(-\omega\, {t}+m\,{\phi})},\quad
 m=\pm0,~\pm1,~\pm2, \cdots,
\label{eq:radial_bath}
\end{equation} 
where $\omega>0$ is the frequency of the acoustic wave and $m$ is the
azimuthal quantum number.  The radial part of this function satisfies
the following differential equation:
\begin{equation}
-\frac{d^2 \psi}{dR_{\text{tort}}^2}+V_\text{eff}(R;\omega,m,\Gamma)\psi=0,
\label{eq:radial_bath}
\end{equation} 
where the tortoise coordinate $R_\text{tort}$ is introduced by
$\partial /\partial R_{\text{tort}}=g(R)\cdot
\partial /\partial R,~g(R)=1-(v^R/c_s)^2$
and the effective potential $V_\text{eff}(R;\omega,m,\Gamma)$ is given
by
\begin{equation}
  V_\text{eff}(R;\omega,m,\Gamma)=-\frac{1}{c_s^2} \left(\omega-\frac{m\,
      v^{\phi}}{R}\right)^2-g(R)\left[\frac{(7-\Gamma)(9+\Gamma)}{4096}
\frac{g(R)}{R^2} 
    -\frac{7-\Gamma}{16R}\frac{dg(R)}{dR}-\frac{m^2}{R^2}\right].
\end{equation}
%
Since $g=0$ at the acoustic horizon $R=R_\text{H}$ and $v^{\phi}$
approaches zero at a point $R_\text{f}$ far from the acoustic horizon, the
asymptotic forms of the effective potential $V_\text{eff}$ are
\begin{equation}
V_\text{eff} \approx
\begin{cases}
-\dfrac{1}{c_s^2}\left(\omega-\dfrac{mv^{\phi}}{R}\right)^2
\qquad&\text{for}\quad R\rightarrow R_\text{H}
\\
-\dfrac{\omega^2}{c_s^2} \qquad&\text{for}\quad R\rightarrow R_\text{f}.
\end{cases}
\label{eq:veff_asymp}
\end{equation}
Therefore, the WKB solutions of Eq.~\eqref{eq:radial_bath}  are
\begin{equation}
\psi=
\begin{cases}
{\displaystyle \exp\left(-i{\int^{R_\text{tort}}}\dfrac{1}{c_s}
\left(\omega-\frac{mv^{\phi}}{R}\right) dR_\text{tort}\right)}
&\quad \text{for}\quad R\sim R_\text{H}\\
{\displaystyle 
C_\text{in}\,\exp\left(-i\int^{R_\text{tort}}\dfrac{\omega}{c_s}dR_\text{tort}
\right) +C_\text{out}\,\exp\left(i\int^{R_\text{tort}} 
\dfrac{\omega}{c_s}dR_\text{tort}\right)}
    &\quad \text{for}\quad R \sim R_\text{f}.
\end{cases}
\label{eq:asympt}
\end{equation}
We take a purely ingoing solution near the acoustic horizon.  From the
conservation of the Wronskian
$W=\psi^{*} (d\psi/dR_\text{tort}) -\psi(d\psi^{*}/dR_\text{tort}$) at
$R=R_\text{H}$ and $R=R_\text{f}$, we obtain the relation between the
reflection and the transmission rates of the acoustic wave:
\begin{equation}
\left|\frac{C_\text{out}}{C_\text{in}}\right|^2
+\frac{c_s(R_\text{H})}{c_s(R_\text{f})}\cdot \frac{\omega-m\,\Omega_\text{H}}{\omega} 
\left|\frac{1}{C_\text{in}}\right|^2=1,
\label{eq:reflect_trans}
\end{equation}
where  $\Omega_\text{H}\equiv v^{\phi}(R_\text{H})/R_\text{H}$
is the angular velocity of the background flow at the acoustic horizon. 
From the relation \eqref{eq:reflect_trans}, we see that the reflection rate
$\left|C_\text{out}/C_\text{in}\right|^2$ can exceed unity when the
frequency $\omega$ satisfies the superradiant condition
\begin{equation}
0<\omega < m\,
\Omega_\text{H}=m\frac{v^{\phi}(R_\text{H})}{R_\text{H}}
=\frac{mL}{(R_{\text{H}})^2} \quad \text{with} \quad m\,L>0.
\label{eq:superrad_cond}
\end{equation}

\begin{figure}[H]
\centering
\includegraphics[width=0.8\linewidth]{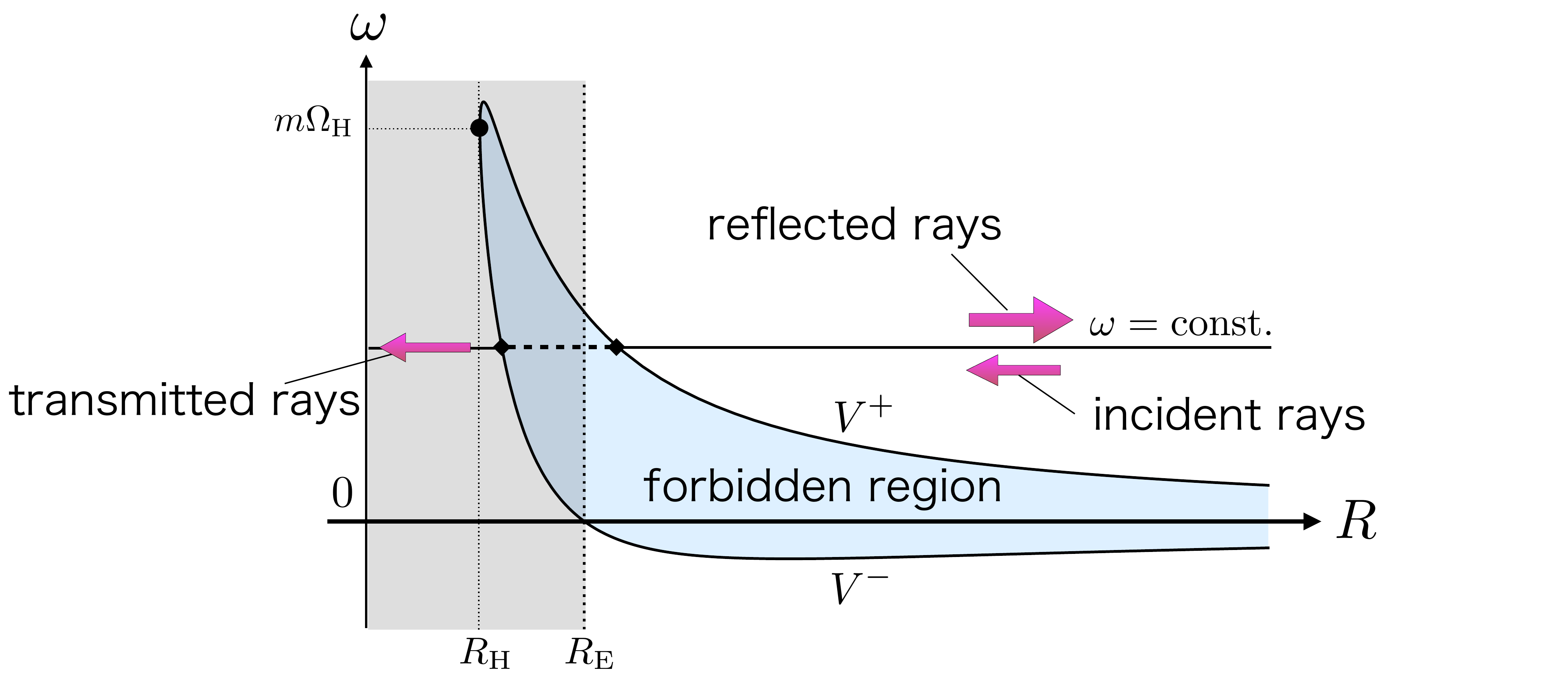}
\caption{The effective potentials $V^{\pm}$ with the parameters
  $\Gamma=4/3, \, c_s(R_s)=1,\, L=1.2R_s$ and $m=20$. The location of the
  acoustic horizon is $R_\text{H}$. The region between the
  acoustic horizon $R_\text{H}$ and the acoustic ergosurface
  $R_\text{E}$ corresponds to the acoustic ergoregion.}
\label{fig:bathtub_veff}
\end{figure}

Now we consider the short wavelength limit of the wave (the
  eikonal limit). Although the amplification factor of
  scattered waves cannot be determined by the analysis with the eikonal
limit, we can discuss the condition of superradiance.  In the
leading order of the eikonal approximation, the phase function $S$ of
the wave function $\delta\Phi \propto e^{iS}$
satisfies the eikonal equation
 \begin{equation}
 \label{eq:HJ}
 s^{\mu\nu}\frac{\partial S}{\partial x^{\mu}}\frac{\partial
   S}{\partial x^{\nu}}
 =0.
\end{equation}
We introduce the acoustic rays as integral curves of the
vector field defined by
\begin{equation}
k^\mu\equiv \frac{dx^{\mu}}{d\lambda}=s^{\mu\nu}\frac{\partial S}{\partial
  x^{\nu}}, 
\label{eq:relation_Ham}
\end{equation}
%
where $\lambda$ is an affine parameter.  The acoustic rays propagate
in the direction perpendicular to the acoustic wave front.  For the
axisymmetric stationary flow, the phase function $S$ can be separated
as $S=-\omega\, {t}+m\,{\phi}+S_{R}(R)$, and Eq.~\eqref{eq:HJ} with the
metric \eqref{eq:bathtub_line_element} yields
\begin{equation}
    s^{tt}\,\omega^2-2s^{t\phi}\,\omega\, m +s^{\phi\phi}\,m^2 
+s^{RR}\left(\frac{dS_R}{dR}\right)^2=0.
\end{equation}
Then using the relation \eqref{eq:relation_Ham}, we obtain 
the equation for the radial component of the acoustic rays,
\begin{equation}
\left(\frac{dR}{d\lambda}\right)^2=-s^{RR}(s^{{t}{t}}\,
\omega^2-2s^{{t}{\phi}}\,\omega\,  m
  +s^{{\phi}{\phi}}\,m^2)=
\frac{1}{c_{s}^2}\,(\omega-V^{+})(\omega-V^{-})
 \geq 0, 
\label{eq:r_geo} 
\end{equation}
where $V^{\pm}$ are  effective potentials for the acoustic rays
defined by
\begin{equation}
  V^{\pm}\equiv m\,\frac{-s_{t{\phi}}\pm 
\sqrt{(s_{t{\phi}})^2-s_{{t}{t}}\,s
_{{\phi}{\phi}}}}{s_{{\phi}{\phi}}}
=m\,\frac{v^{\phi}\pm
  \sqrt{c_{s}^2-(v^{R})^2}}{R}. 
\label{eq:plis_minus}
\end{equation}
From Eq.~\eqref{eq:r_geo}, 
the interval of $R$ determined by $V^{-}(R) \leq \omega \leq V^{+}(R)$ 
is forbidden for the propagation of the acoustic rays
(Fig.~\ref{fig:bathtub_veff}). At
$R=R_\text{H}$, $V^{+}=V^{-}=m\,\Omega_\text{H}$.
\noindent
The interpretation of superradiance in terms of acoustic rays is as
follows: Owing to the  wave effect (i.e., tunneling), the incident rays with
frequency (energy) satisfying the superradiant condition~
\eqref{eq:superrad_cond} can appear in the negative energy state
region by penetrating the potential barrier. The
transmitted rays finally fall into the acoustic horizon. In the course of
this process, the rays reflected by the potential wall are
amplified.


\section{Magnetoacoustic geometry} 
Now, we investigate the effective
geometry defined for the MHD waves in the eikonal limit. To define the
magnetoacoustic metric, we consider the magnetic pressure-dominated
and the gas pressure-dominated cases.  Then, to investigate the structure
of the magnetoacoustic geometry, we prepare an axisymmetric stationary
solution of MHD inflow.


\subsection{Eikonal equation and magnetoacoustic metric}  
The basic equations for the ideal MHD are
\begin{align}
&\frac{\partial \rho}{\partial t}+\nabla\cdot(\rho\, \bm{v
})=0,
\label{eq:cont}\\
&\nabla \cdot \bm{B}=0,
\label{eq:divb}\\
&\frac{\partial \bm{B}}{\partial t}=\nabla \times (\bm{v} \times \bm{B}),
\label{eq:induc}\\
&\frac{\partial \bm{v}}{\partial t}+(\bm{v} \cdot
  \nabla)\bm{v}+\frac{\nabla p}{\rho}+\frac{1}{4\pi \rho}\bm{B} 
\times (\nabla \times \bm{B})+\bm{F}^{\text{MHD}}_\text{ex}=0,
\label{eq:euler}
\end{align}
where $\bm{B}$ is the magnetic field and
$\bm{F}^{\text{MHD}}_\text{ex}$ is an external force to control the
MHD flow. The barotropic equation of state $p=p(\rho)$ is also
  assumed. 
We decompose all quantities to the stationary background part and 
the perturbed part.  Assuming that the wavelength of
  perturbations is small compared to the scale of spatial variation of
the background quantities, the equations of the first-order
perturbations are written as follows:
\begin{align}
&\frac{D\delta \rho}{D t}+\rho\nabla\cdot  \delta\bm{v
}=0,
\label{eq:cont1}\\
&\nabla \cdot \delta \bm{B}=0,
\label{eq:divb1}\\
&\frac{D \delta \bm{B}}{D t}=\nabla \times (\delta \bm{v} \times \bm{B}),
\label{eq:induc1}\\
&\frac{D \delta \bm{v}}{D t}+\frac{c_{s}^2}{\rho}\nabla \delta
  \rho+\frac{1}{4\pi \rho}\bm{B} \times (\nabla \times \delta
  \bm{B})=0,
\label{eq:euler1}
\end{align}
where $\bm{v}, \bm{B}, \rho$ are background quantities and the
Lagrange derivative is introduced by
$D/Dt=\partial/\partial t+\bm{v}\cdot\nabla$.  Then, applying
$D/Dt$ to Eq. \eqref{eq:euler1} and using Eqs. \eqref{eq:cont1} and
\eqref{eq:induc1}, we obtain the wave equation for $\delta \bm{v}$,
\begin{equation}
\frac{D^2 \delta \bm{v} }{D t^2}-c_{s}^2\,\nabla (\nabla \cdot \delta
\bm{v})+\bm{V}_\text{A} \times  \nabla \times \left(\nabla \times (\delta
    \bm{v} \times \bm{V}_\text{A})\right)=0,
\label{eq:perturbed_eq}
\end{equation}
where $\bm{V}_\text{A}=\bm{B}/\sqrt{4\pi \rho}$ is the
Alfv\'en velocity, which represents the propagating velocity of the 
transverse MHD wave mode along the magnetic field lines. 

Hereafter, we consider two-dimensional flows in the
$xy$ plane and assume that all the vectors have no $z$
  components.  By applying the Helmholtz theorem, the velocity
    perturbation can be described by using two scalar functions $\Phi,
    \Psi$ as 
\begin{equation}
\delta \bm{v}=
\begin{pmatrix} \partial_{x} \Phi\\ \partial_{y}\Phi \\ \end{pmatrix}
 +\begin{pmatrix}\partial_{y}\Psi \\ -\partial_{x} \Psi \\ \end{pmatrix}
 =\begin{pmatrix}\partial_{x} & \partial_{y} \\ \partial_{y}&
  -\partial_{x}\\ \end{pmatrix}
\begin{pmatrix} \Phi \\ \Psi \\ \end{pmatrix}.
\label{eq:decompose}
\end{equation}
  Substituting this
into Eq.~\eqref{eq:perturbed_eq}, we obtain
\begin{equation}
 \frac{D^2}{Dt^2}
\begin{pmatrix} \Phi \\ \Psi\\ \end{pmatrix}
=\left[\begin{pmatrix} c_{s}^2\,\nabla^2 & 0 \\ 0&0 \\ \end{pmatrix}+
 \begin{pmatrix} \hat{L}_{1}^2 & \hat{L}_{1}\hat{L}_{2} \\ \hat{L}_{1}\hat{L}_{2}& \hat{L}_{2}^2 \end{pmatrix}
  \right]
\begin{pmatrix}\Phi \\ \Psi \\ \end{pmatrix},
\label{eq:phi_psi}
\end{equation}
where $\hat{L}_{1}$ and $\hat{L}_{2}$ are the following derivative operators:
\begin{equation}
\hat{L}_{1}=(V_\text{A})_y\,\partial_{x}-(V_\text{A})_x\,\partial_{y} ,\quad
 \hat{L}_{2}=(V_\text{A})_x\,\partial_{x}+(V_\text{A})_y\,\partial_{y},
\end{equation}
and they are approximately commutable with each other since the
  spatial derivatives of the background quantities are assumed to be small.
We define the magnetoacoustic metric as the components of the eikonal
equation for the MHD waves.  We substitute the following form of the
wave function into Eq.~\eqref{eq:phi_psi}:
\begin{equation}
 \Phi =|\Phi|\,e^{iS},\quad \Psi= |\Psi|\,e^{iS},
\end{equation} 
where $S$ is the phase of the wave functions.  Up to the
leading order of the eikonal approximation, we obtain
\begin{equation}
\left(\frac{\partial S}{\partial t}+\bm{v} \cdot \nabla S\right)^2
\begin{pmatrix} {|\Phi|} \\ {|\Psi|}\\ \end{pmatrix}
=\left[\begin{pmatrix} c_{s}^2\,|\nabla S|^2 & 0 \\ 0&0 \\ \end{pmatrix}+
 \begin{pmatrix} (\hat{L}_{1}S)^2 & (\hat{L}_{1}S)(\hat{L}_{2} S)\\ (\hat{L}_{1}S)(\hat{L}_{2}S)& (\hat{L}_{2}S)^2 \end{pmatrix}
  \right]
\begin{pmatrix}{|\Phi|} \\ {|\Psi|} \\ \end{pmatrix}.
\label{eq:phi_psi_1}
\end{equation}
For Eq.~\eqref{eq:phi_psi_1} to have nontrivial solutions, 
we obtain the following eikonal equation
\footnote{As Eq.~\eqref{eq:eik} is the quartic form for 
$\partial S/\partial x^{\mu}$, it can be written as
\begin{equation}
\nonumber {{M}}^{\mu\nu\lambda \sigma}\frac{\partial S}{\partial x^{\mu}} \frac{\partial S}{\partial x^{\nu}}\frac{\partial S}{\partial x^{\lambda}}\frac{\partial S}{\partial x^{\sigma}}=0,
\label{eq:Finsler}
\end{equation}
where $M^{\mu\nu\lambda \sigma}$ is the coefficient of the eikonal
equation. If we define a function $F(x,y)$ as
$F(x,y)=\left({{M}}^{\mu\nu\lambda \sigma}{y_{\mu} }{y_{\nu}
    }{y_{\lambda}}{y_{\sigma}}\right)^{1/4}$
with $y_{\mu}=\partial_{\mu}S$, we see that this function satisfies
the condition $F(x,\alpha y)=\alpha F(x,y)$.  This kind of function
defines a distance $ds=F(x,dx)$ of the Finsler geometry \cite{Bao2000},
which is a generalization of the Riemannian geometry.  Since the null
condition $ds=0$ corresponds to the  eikonal equation \eqref{eq:eik}, the motion
of MHD waves may provide an effective Finsler geometry.}
:
 \begin{equation}
\left(\frac{\partial S}{\partial t}+\bm{v}\cdot \nabla  S\right)^4-(c_{s}^2+V_\text{A}^2)|\nabla S|^2\left(\frac{\partial S}{\partial t}+\bm{v}\cdot \nabla  S\right)^2
+(c_{s}^2\,|\nabla S|^2)(\bm{V}_{\text{A}} \cdot \nabla S)^2=0,
\label{eq:eik}
\end{equation}
where we used relations $(\hat L_1 S)^2+(\hat L_2
S)^2=V_\text{A}^2\,|\nabla S|^2$  
and $\hat L_2 S = \bm{V}_\text{A} \cdot \nabla S$. 
We rewrite the eikonal equation 
as
\begin{equation}
\left(\frac{\partial S}{\partial t}+\bm{v}\cdot \nabla  S\right)^2=
\frac{V_\text{M}^2\,|\nabla S|^2}{2}\left[1 \pm 
\sqrt{1-4\left(\frac{c_{s}\,V_\text{A}}{V_\text{M}^2}\right)^2\,
\left(\frac{\bm{b} \cdot \nabla S}{|\nabla S|}\right)^2}\,\right],
\label{eq:squareroot}
\end{equation}
%
where $V_\text{M} \equiv \sqrt{c_{s}^2+V_\text{A}^2}$, and
$\bm{b}\equiv \bm{V}_\text{A}/V_\text{A}$ represents the direction of
background magnetic field lines.  The plus and minus signs correspond
to the fast and the slow magnetoacoustic wave modes,
respectively. Note that there is no transverse Alfv\'en mode in our
analysis, as we are considering two-dimensional flows.  Since the
eikonal equation \eqref{eq:squareroot} does not have the
quadratic form with respect to the derivative term of the phase $S$,
it is not possible to introduce the quadratic (Riemannian) acoustic
metric.  In order to circumvent this problem, we introduce a parameter
  \begin{equation}
  \eta \equiv \left(\frac{c_sV_\text{A}}{V_\text{M}^2}\right)^2.       
  \end{equation}
 If we regard $\eta$ as a small parameter, it is
  possible to expand the square root term in 
  Eq.~\eqref{eq:squareroot} as
\begin{equation}
\sqrt{1-4\eta \left(\frac{\bm{b} \cdot \nabla S}{|\nabla S|}\right)^2}
=
1-2 \eta \left(\frac{\bm{b} \cdot \nabla S}{|\nabla S|}\right)^2
+O\left(\eta^2 \right).
\label{eq:expansion}
\end{equation}
%
 The condition $\eta\ll1$ can be satisfied for
$(c_s/V_\text{A})^2\ll1$ or $(V_\text{A}/c_s)^2\ll1 $ because the value of
  the parameter is
\begin{equation}
\eta\approx
\begin{cases}
            (c_s/V_\text{A})^2\ll1\qquad
         &\text{(magnetic pressure-dominated case)}\\
       (V_\text{A}/c_s)^2\ll1
\qquad
         &\text{(gas pressure-dominated case)}.
\end{cases}
\end{equation}
%
Then Eq.~\eqref{eq:squareroot} 
becomes 
\begin{equation}
 \left(\frac{\partial  S}{\partial t}+\bm{v}\cdot \nabla  S\right)^2
\approx
\begin{cases}
V_\text{M}^2\left(|\nabla S|^2-\eta\, (\bm{b}\cdot \nabla S)^2\right)
\quad &\text{(fast mode)}\\
\eta\, V_\text{M}^2\,(\bm{b}\cdot \nabla S)^2\quad &\text{(slow mode)}
\end{cases}.
\label{eq:two_eikonal_eqs}
\end{equation}
Note that in both the magnetic pressure-dominated and 
          gas pressure-dominated cases, the fast wave mode propagates almost 
          isotropically like sound waves.
This is because the right-hand side of the eikonal equation for the fast mode 
consists of the isotropic term $V_\text{M}^2 |\nabla S|^2$ and the small correction 
term $\eta V_\text{M}^2 (\bm{b}\cdot \nabla S)^2$, which 
represents anisotropic effects due to the magnetic field. 
By assigning the coefficients of these equations to the components of
matrices $M_{\text{fast}}^{\mu\nu}$ and $M_{\text{slow}}^{\mu\nu}$,
Eqs. \eqref{eq:two_eikonal_eqs} can be written as
\begin{align}
{M}_{\text{fast}}^{\mu\nu}\,\frac{\partial S}{\partial x^{\mu}}\frac{\partial S}{\partial x^{\nu}}=0 \quad&\quad\text{(fast mode)},
\label{eq:eikonal_f} \\ 
M_{\text{slow}}^{\mu\nu}\,\frac{\partial S}{\partial
  x^{\mu}}\frac{\partial S}{\partial x^{\nu}}=0 \quad&\quad\text{(slow mode)},
  \label{eq:eikonal_s}
\end{align}
where ${M}_{\text{fast}}^{\mu\nu}$ and ${M}_{\text{slow}}^{\mu\nu}$ are
\begin{equation} 
{M}_{\text{fast}}^{\mu\nu}=
\begin{pmatrix}
 -1 & -v^i \\ -v^i & \ V_\text{M}^2\,\delta^{ij}-\left(v^i\,v^j+\eta\,
 V_\text{M}^2\, b^i b^j\right)
\end{pmatrix},\ \ {M}_{\text{slow}}^{\mu\nu}=\begin{pmatrix}
 -1 & -v^i \\ -v^i & \ -\left(v^i\,v^j-\eta\, V_\text{M}^2\, b^i b^j\right)
\end{pmatrix},\ i,j=1,2.
\label{eq:s_fast_slow_comp}
\end{equation}
The matrix  ${M}_{\text{fast}}^{\mu\nu}$ has an inverse, 
and we obtain the magnetoacoustic line element 
$ds_\text{fast}^2=({M}_{\text{fast}})_{\mu\nu}\,dx^\mu dx^\nu$ as
\begin{equation}
 ds_\text{fast}^2\propto
-\left[(V_\text{M}^2- \bm{v}^2)-\eta \, (\bm{b}\cdot \bm{v})^2\right]dt^2
 -2\left[ v_i +\eta \, b_{i}(\bm{b}\cdot \bm{v})\right]dt\,dx^{i}
  +\left( \delta_{ij}+\eta \, b_i b_j\right)dx^{i}dx^{j}.
\label{eq:acmetric_fast}
\end{equation}
We write this metric in the polar coordinates $(R, \phi)$ and 
apply the following coordinate transformation:
\begin{equation}
dt \rightarrow dt+\frac{ v^R}{V_\text{M}^2-\eta\,
  V_\text{M}^2(b^{R})^2-(v^R)^2}dR,\quad
d\phi \rightarrow d\phi+\frac{ v^R\, v^\phi+\eta\, b^R\, b^\phi\, V_\text{M}^2}
{V_\text{M}^2-\eta\, V_\text{M}^2\,(b^{R})^2-(v^R)^2}\frac{dR}{R}.
\label{eq:coordinate_tr}
\end{equation}
Then the magnetoacoustic metric \eqref{eq:acmetric_fast} can be written as
\begin{equation}
\begin{aligned}
     ds_\text{fast}^2\propto
&-\left[(V_\text{M}^2- \bm{v}^2)-\eta \, (\bm{b}\cdot \bm{v})^2\right]d{t}^2
-2\left[ v^{\phi} +\eta\, b^{\phi}(\bm{b}\cdot \bm{v})\right]R\, d{t}\,d{\phi}
\\ 
 &\qquad\qquad
+\frac{dR^2}{1-\eta\,
  (b^{R})^2-(v^R/V_\text{M})^2}  +\left[1+\eta\, (b^{\phi})^2\right]R^2d{\phi}^2,
  \label{eq:acmetric_fast_tilde}
\end{aligned}
\end{equation}
where $t$ and $\phi$ are new coordinates after the coordinate
transformation \eqref{eq:coordinate_tr}.  As in the
case of the acoustic black holes for perfect fluids,  the
effective horizon and ergosurface for MHD waves (magnetoacoustic
horizon and ergosurface)  are defined as points where the
following conditions hold:
\begin{align}
 \label{eq:Mag_hor} (v^{R})^2&=V_\text{M}^2-\eta V_\text{M}^2 (b^{R})^2
&  & \text{(magnetoacoustic horizon)}, \\ 
  \bm{v}^2&=V_\text{M}^2-\eta (\bm{b} \cdot \bm{v})^2
&  & \text{(magnetoacoustic ergosurface)}.
\label{eq:Mag_ergo}
\end{align}
{Note that these conditions can be obtained without the 
coordinate transformation \eqref{eq:coordinate_tr}, as well 
by following the definition of the 
black hole horizon and the ergosurface in general relativity. 
Namely, we can derive the horizon condition \eqref{eq:Mag_hor} 
by searching the surface where a killing vector 
$\xi\equiv \xi_{(t)}+(v^\phi/R)\xi_{(\phi)}$ becomes null. 
The condition for ergosurface \eqref{eq:Mag_ergo} 
is given by $\xi_{(t)}^2=0$.}
According to the eikonal equations \eqref{eq:two_eikonal_eqs}, the
fast mode propagating in the radial direction
($\nabla S \propto \bm{e}_{R}$) has the propagating velocity
$V_\text{M}^2-\eta\, V_\text{M}^2\,(b^{R})^2$. Hence, the
magnetoacoustic horizon defined by relation \eqref{eq:Mag_hor} is the
one-way boundary for the fast mode propagation.  While inside of the
magnetoacoustic ergoregion defined by
$\boldsymbol{v}^2\ge V_\text{M}^2-\eta\,(\boldsymbol{b}\cdot\boldsymbol{v})^2$,
the fast mode cannot propagate in the counter direction of the
background flow velocity $\bm{v}$.

For the slow mode, on the other hand, it is not possible to introduce the 
magnetoacoustic metric because the matrix $M_{\text{slow}}^{\mu\nu}$
does not have an inverse due to the lack of the isotropic term
$\propto|\nabla S|^2=(\partial_{x}S)^2+(\partial_{y}S)^2$,
and the inner product between vectors cannot be defined.
We can rewrite the eikonal equation \eqref{eq:eikonal_s} for the slow mode as
\begin{equation}
\frac{\partial  S}{\partial t}+\bm{v}_{\pm}\cdot \nabla S=0, 
\label{eq:slow_eikonal}
\end{equation}
where
$\bm{v}_{\pm}\equiv \bm{v}\pm (c_sV_\text{A}/V_\text{M}) \bm{b}$.
Since the eikonal equation \eqref{eq:slow_eikonal} is the advective
type, we see that the propagation is restricted along $\bm{v}_{\pm}$,
and plus and minus signs correspond to outgoing and ingoing
  waves, respectively.  Although the magnetoacoustic line element is
not defined, it can be possible to discuss the motion of rays of the
slow mode.


\subsection{Background MHD flow and magnetoacoustic black holes}  
As the background MHD flow, we consider a stationary and axisymmetric
inflow. We assume that the equation of state is polytropic
$p\propto \rho^{\Gamma}$ with $0 \leq \Gamma \leq 4$.
\footnote{In the MHD case, we can consider $\Gamma=0$ 
     ($c_s=0$), and the magnetoacoustic geometries belong to the class on the
     $\alpha=0$ axis in Fig.~\ref{fig:fast_result}. 
   The upper bound $\Gamma \leq 4$ is determined by the 
   condition $\eta \ll 1$. According to Eq.~\eqref{eq:ratio_cs_va}, 
    for $\Gamma >4$, the condition $\eta \ll 1$ is not satisfied for small
    $R$ in the background flow.
} 
To obtain the
background MHD flow, we basically follow Weber and Davis
\cite{Weber1967} who derived the stationary and axisymmetric solution
of the ideal MHD flow. The difference between our treatment and theirs
is that we assume a profile of the radial fluid velocity
$v^{R}(R)$ by introducing an appropriate external force. 
The basic equations for the background flow are
\begin{align}
&{\nabla}\cdot (\rho\, \bm{v})=0,
\label{eq:cont}\\
&\nabla \cdot \bm{B}=0,
\label{eq:divb}\\
&\nabla \times (\bm{v} \times \bm{B})=0,
\label{eq:induc}\\
&(\bm{v} \cdot \nabla)\bm{v}+\frac{\nabla p}{\rho}+\frac{1}{4\pi \rho}\bm{B} 
\times (\nabla \times \bm{B})+\bm{F}^{\text{MHD}}_\text{ex}=0.
\label{eq:euler}
\end{align}
Since the flow is axisymmetric, the components of $\bm{v}$
and $\bm{B}$ depend only on $R$ as
\begin{equation}
\bm{v}=v^{R}(R)\, \bm{e}_{\hat{R}}+v^{\phi}(R)\,\bm{e}_{\hat{\phi}} ,\quad
 \bm{B}=B^{R}(R)\, \bm{e}_{\hat{R}}+B^{\phi}(R)\,\bm{e}_{\hat{\phi}}.
\end{equation}
Equations (\ref{eq:cont}) and (\ref{eq:divb}) are solved as
\begin{equation}
\rho\, v^{R} R = \text{const}\ <0\  ,\quad R B^{R}=\text{const} \ >0.
\label{eq:concon}
\end{equation}
The azimuthal component of the induction equation (\ref{eq:induc}) yields
\begin{equation}
\frac{d}{dR}[v^{\phi}B^{R}-v^{R}B^{\phi}]=0,
\end{equation}
and we obtain the following relation:
\begin{equation}
v^{R}B^{\phi}-v^{\phi}B^{R}=\text{const}\equiv -\Omega_{\text{F}}\,
R\, B^{R}, \label{eq:omega}
\end{equation}
where we used Eq.~\eqref{eq:concon} to define the conserved quantity
$\Omega_{\text{F}}$, which is the angular velocity of the magnetic
field line.  In addition to this quantity, the conservation of the
angular momentum of the fluid $L$ is also derived from the azimuthal
component of Eq.~(\ref{eq:euler}) as
\begin{equation}
R\,v^{\phi}-\frac{B^{R}}{4 \pi \rho\, v^{R}}\ RB^{\phi}=\text{const}\equiv L.
\label{eq:phi_euler}
\end{equation}
From Eqs.~(\ref{eq:omega}) and (\ref{eq:phi_euler}), the azimuthal component
of the fluid velocity $v^{\phi}$ is obtained as a function of 
the radial components $v^{R}$ and $B^{R}$:
\begin{equation}
v^{\phi}=\Omega_{\text{F}}\, R\, \frac{M_\text{A}^2\, L\, R^{-2}\,\Omega_{\text{F}}^{-1}-1}{M_\text{A}^2-1},
\label{eq:vphi}
\end{equation}
where we introduced the radial Alfv\'enic Mach number 
$M_\text{A}^2(R)\equiv 4\pi \rho (v^{R})^2/(B^{R})^2=(v^{R}/V_\text{A}^{R})^2$.
To construct the solution, we require that the background inflow is
the trans-magnetosonic flow.  Such a flow should pass through the
radial Alfv\'en point $R=R_{*}$ where $M_{\text{A}}^2=1$ and the
denominator in Eq.~\eqref{eq:vphi} becomes zero. Hence, the numerator
should go to zero simultaneously to keep the azimuthal component
$v^{\phi}$ finite.  Therefore, the relation between the conserved
quantities $L$ and $\Omega_{\text{F}}$ is obtained as
\begin{equation}
L=\Omega_{\text{F}} R_{*}^2.
\label{eq:L_and_Omega}
\end{equation}
Taking this condition into account, $v^{\phi}$ can be reduced to
\begin{equation}
v^{\phi}=\frac{\Omega_{\text{F}}\, R_{*}}{v_{*}^{R}}\frac{v^{R}-(R/R_{*})v_{*}^{R}}{M_{\text{A}}^2-1},
\label{eq:vphi2}
\end{equation}
where $v_{*}^{R}=v^{R}(R_{*})$.  In the same way, the azimuthal
component of the magnetic field is obtained from Eqs.~\eqref{eq:omega}
and \eqref{eq:vphi} with the relation \eqref{eq:L_and_Omega},
\begin{equation}
B^{\phi}=
-\,\frac{B^{R}\,\Omega_{\text{F}}}{v^{R}_{*}R_{*}}\frac{R^2-{R_{*}}^2}{M_{\text{A}}^2-1}.
\label{eq:Bphi}
\end{equation}

\begin{figure}[H]
\centering
\includegraphics[width=0.6\linewidth]{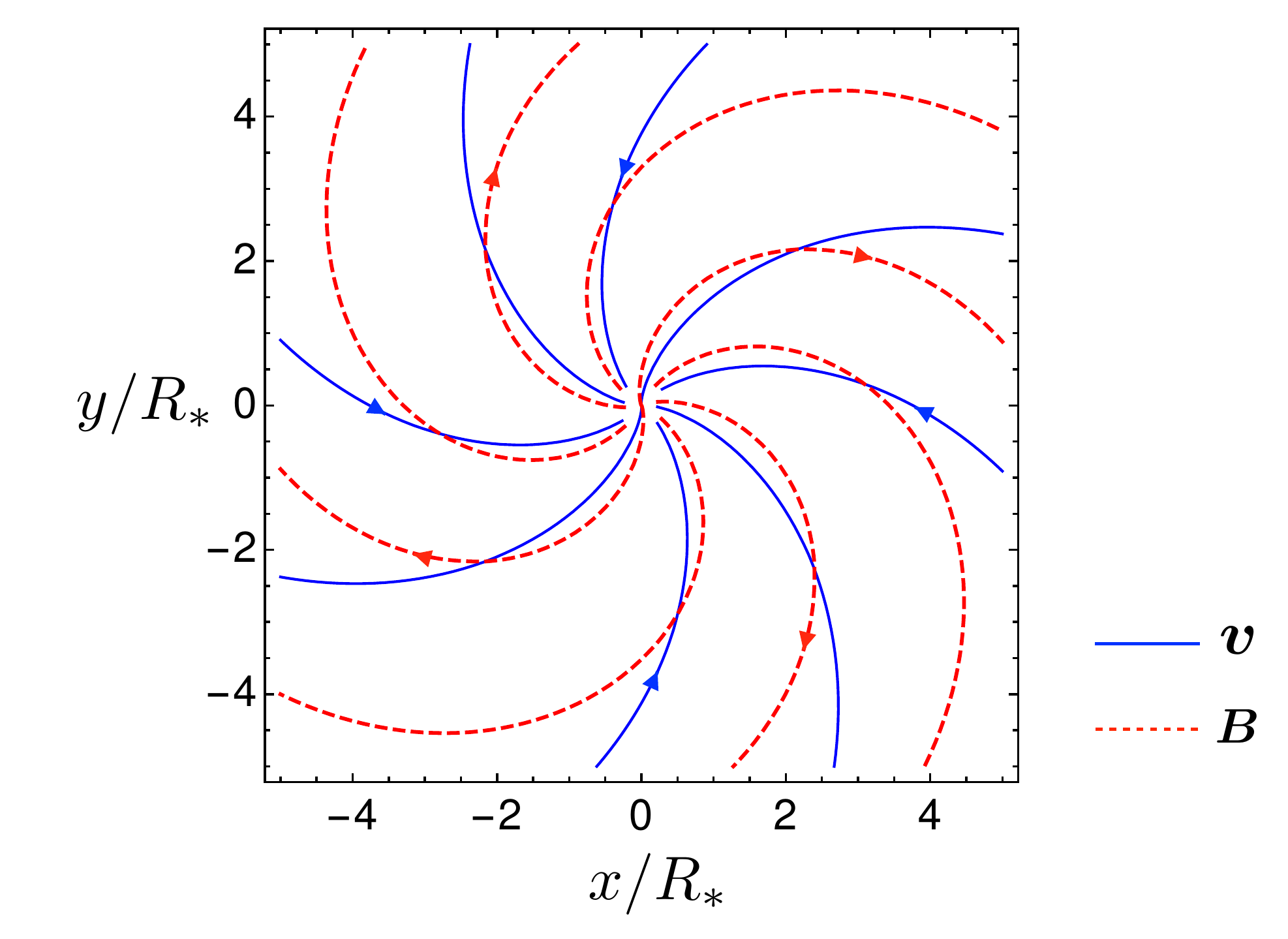}  
\caption{Background MHD inflow with $\Gamma=4/3$ and $\beta=0.08$.
  The solid arrows (blue) are the stream lines of velocity $\bm{v}$,
  and the dotted arrows (red) represent the magnetic field lines of
  $\bm{B}$. The shapes of stream lines and magnetic field lines 
  depend only on values of the parameter $\beta$.}
\label{fig:flow}
\end{figure}

To obtain the background quantities as functions of $R$, we first
consider the radial components $v^{R}$ and $B^{R}$.  From
Eq.~\eqref{eq:divb}, the radial component of the magnetic field is
\begin{equation}
B^{R}=B_{*}^{R}\frac{R_{*}}{R},
\label{eq:Br}
\end{equation}
where $B_{*}^{R}\equiv B^{R}(R_{*}) >0$.  The radial component of
Eq.~\eqref{eq:euler} is
\begin{equation}
v^{R}\frac{d v^{R}}{dR}-\frac{v^{\phi}}{R}+\frac{1}{\rho}\frac{dp}{dR}
+\frac{1}{4\pi \rho}\frac{B^{\phi}}{R}\frac{d}{dR}(R B^{\phi})+F^{\text{MHD}}_\text{ex}=0.
\label{eq:euler_r}
\end{equation}
This equation determines $v^R(R)$ because $v^{\phi}$ and
$B^\phi$ are already expressed by $v^R$ from \eqref{eq:vphi2} and
\eqref{eq:Bphi}, and the pressure $p=p(\rho)$ can be written with
  $v^{R}$ through the polytropic equation of state and the
  conservation law \eqref{eq:concon}. We can assume an
  arbitrary profile of $v^{R}$ by choosing
  $F^{\text{MHD}}_\text{ex}(R)$ to satisfy Eq.~\eqref{eq:euler_r}.
We choose the draining bathtub-type inflow, 
which behaves as
\footnote{Writing the radial velocity as 
$v^R \propto (R/R_{*})^{-\sigma}$, we see that the 
same classification of the magnetoacoustic geometry is obtained for $0<\sigma <1$.  
In this paper, we choose $\sigma =1/2$ to make calculations simpler.}
\begin{equation}
v^{R}=v_{*}^{R}\left(\frac{R}{R_{*}}\right)^{-{1}/{2}}, 
\label{eq:draining}
\end{equation}
where $v_{*}^{R}\equiv v^{R}(R_{*}) <0$ is the radial fluid velocity
at the radial Alfv\'en point. 
This profile is different from the standard bathtub model $v^R \propto 1/R$. 
The reason why we do not take the standard one is that for $v^R \propto 1/R$, 
the radial Alfv\'enic Mach number $M_\text{A}$ becomes constant and the radial 
Alfv\'en point does not exist.
Substituting $v^R$ [Eq.~\eqref{eq:draining}]
into Eqs.~\eqref{eq:vphi2} and \eqref{eq:Bphi}, we obtain the azimuthal 
components $v^{\phi}$ and $B^{\phi}$, 
\begin{eqnarray}
&&v^{\phi}=-\beta\, v_{*}^{R}\left[\left(\frac{R}{R_{*}}\right)^{{1}/{2}}+\left(\frac{R}{R_{*}}\right)^{-{1}/{2}}+1\right],
\label{eq:mhd_radial}\\
\nonumber \\
&&  B^{\phi}=-\beta B_{*}^{R}  \left[1+\left(\frac{R}{R_{*}}\right)^{{1}/{2}}\right]\left[1+\left(\frac{R}{R_{*}}\right)^{-1}\right],
\label{eq:mhd_phi}
\end{eqnarray}
where we introduced a parameter
\begin{equation}
\beta \equiv \Omega_{\text{F}} R_{*}/v_{*}^{R}.
\end{equation}
This constant represents the ratio of the angular velocity of the
magnetic field lines to the radial component of the fluid velocity
$v^{R}$ at the radial Alfv\'en point $R_{*}$. The density profile
$\rho(R)$ is also obtained from Eq.~\eqref{eq:concon}.  Although the 
obtained flow has a singularity at $R=0$ and the magnetic field is not
divergence free there, we evaluate the properties of the MHD waves 
in the region, except for this singularity.
Using the quantities of
the background flow, we obtain the sound velocity ${c}_{s}$ and  Alfv\'en
velocity ${V}_{\text{A}}$  as functions of $R/R_{*}$:
\begin{align}
&{c}_{s}^2 =\alpha^2 (v_{*}^{R})^2\left(\frac{R}{R_{*}}\right)^{-(\Gamma-1)/2}, \label{eq:cs}\\
&
{V}_{\text{A}}^2 =(v_{*}^{R})^2 \left\{\left(\frac{R}{R_{*}}\right)^{-3/2}+\beta^2\left(\frac{R}{R_{*}}\right)^{1/2} \left[1+\left(\frac{R}{R_{*}}\right)^{{1}/{2}}\right]^2\left[1+\left(\frac{R}{R_{*}}\right)^{-1}\right]^2\right\},
\label{eq:va}
\end{align}
where we introduced another parameter,
\begin{equation}
\alpha \equiv -c_{s*}/v^{R}_{*}~>0,
\end{equation}
which is the ratio of the sound velocity to the radial component 
of the flow velocity $v^{R}$ at the radial Alfv\'en point $R_{*}$.
The structure of the MHD flow obtained here is shown 
in Fig.~\ref{fig:flow}. Note that our MHD flow is not reduced to
 the draining bathtub model Eq.~\eqref{eq:bathtub} for any values of
$(\alpha,\beta)$. This is because the condition \eqref{eq:L_and_Omega} 
is imposed via the relation between
$L$ and $\Omega_{\text{F}}$ in this magnetized fluid system. Hence, our
two-dimensional MHD flow is not just the generalization of the
draining bathtub model for the perfect fluids. 

\begin{figure}[H]
   \begin{center}
   \includegraphics[width=0.7\linewidth]{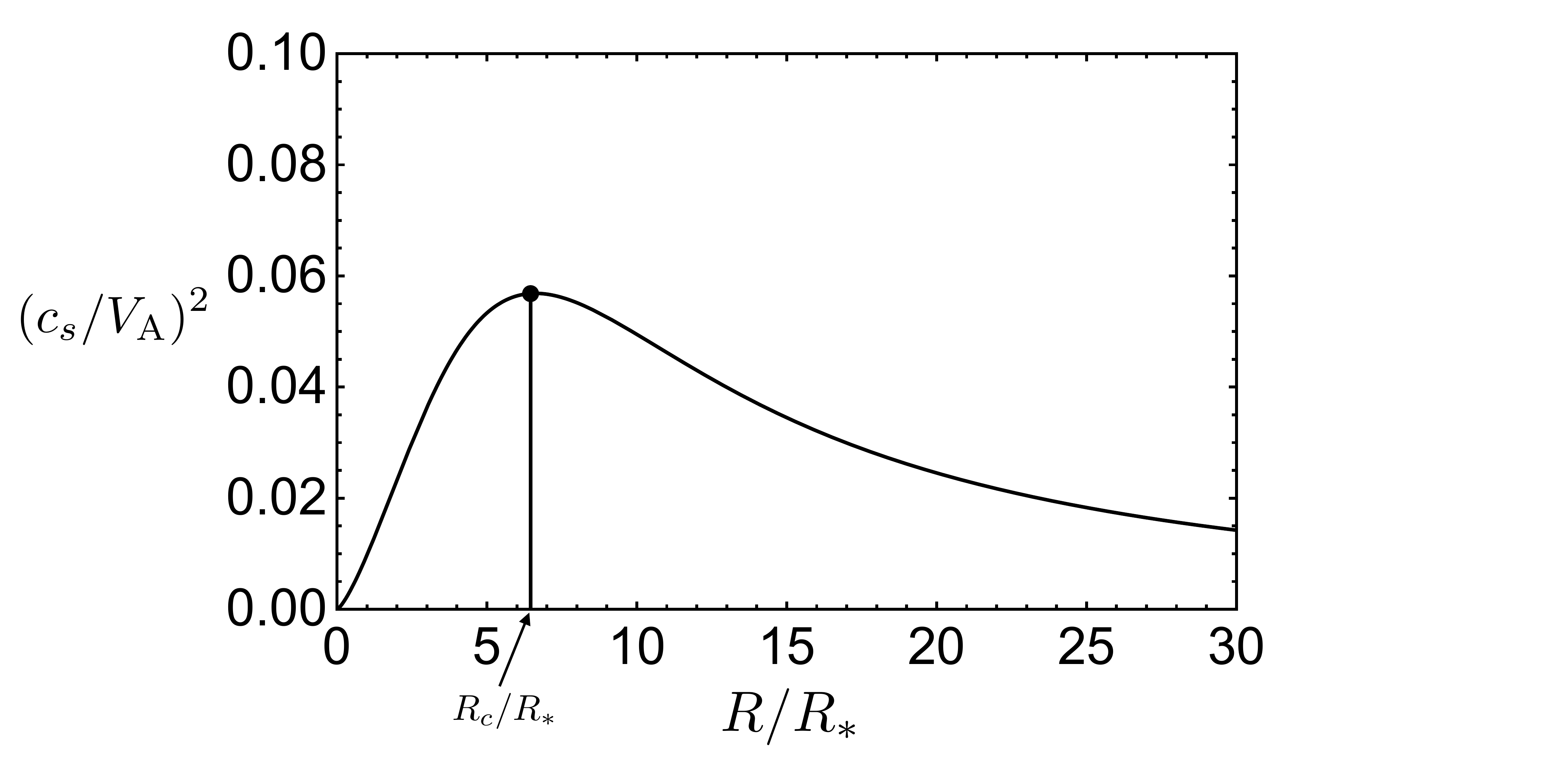}
  \end{center}
 \caption{$R$ dependence of $(c_s/V_{\text{A}})^2$
      for $\Gamma=4/3$ and $(\alpha,\beta)=(0.1, 0.04)$.
     Note that $(c_s/V_\text{A})^2$ becomes maximum at $R=R_c$.}
  \label{fig:VA_cs}
\end{figure}

In order to apply this background flow to the magnetoacoustic metric 
\eqref{eq:acmetric_fast_tilde}, the following condition must be satisfied:
\begin{equation}
\eta(R)=\left(\frac{c_sV_\text{A}}{V_\text{M}^2}\right)^2
\ll 1 \quad \text{for all} \ R.
\label{eq:eta-cond}
\end{equation}
Since $\eta=\left(c_s V_\text{A}/V_\text{M}^2\right)^2 \ll 1$ holds
for both the $(V_\text{A}/ c_s)^2 \ll1$ and $(c_s/V_\text{A})^2\ll1$
cases, we need to check this condition for our background MHD flow.
From \eqref{eq:cs} and \eqref{eq:va}, the ratios $(c_s/V_\text{A})^2$
at $R\rightarrow 0$ and $R\rightarrow \infty$ are
\begin{equation}
\left(\frac{c_s}{V_\text{A}}\right)^2 \xrightarrow{R \rightarrow 0}
\frac{\alpha^2}{1+\beta^2}\left(\frac{R}{R_{*}}\right)^{(4-\Gamma)/2}, \quad
\left(\frac{c_s}{V_\text{A}}\right)^2 \xrightarrow{R \rightarrow \infty} 
\frac{\alpha^2}{\beta^2}\left(\frac{R}{R_*}\right)^{-(\Gamma+2)/2}.
\label{eq:ratio_cs_va}
\end{equation}
For the polytropic index $0\leq \Gamma \leq 4$, these values
approach zero, and only
the magnetic pressure-dominated case $\left(c_s/V_\text{A}\right)^2 \ll 1$
can be realized in our model (see Fig.~\ref{fig:VA_cs}). As the criterion for the condition
\eqref{eq:eta-cond}, we use the maximum value
$\eta_c\equiv \left.\left(c_s/V_\text{A}\right)^2\right|_{R=R_c}$
where $R_c$ is a maximum point of $(c_s/V_\text{A})^2$.


\section{classification of the magnetoacoustic geometry}
Now, we discuss the magnetoacoustic geometry introduced in the
previous section.  As a probe of the geometry, we investigate the
motion of the magnetoacoustic rays. 

\subsection{Fast mode}

\begin{figure}[H]
\centering
\includegraphics[width=1.05\linewidth]{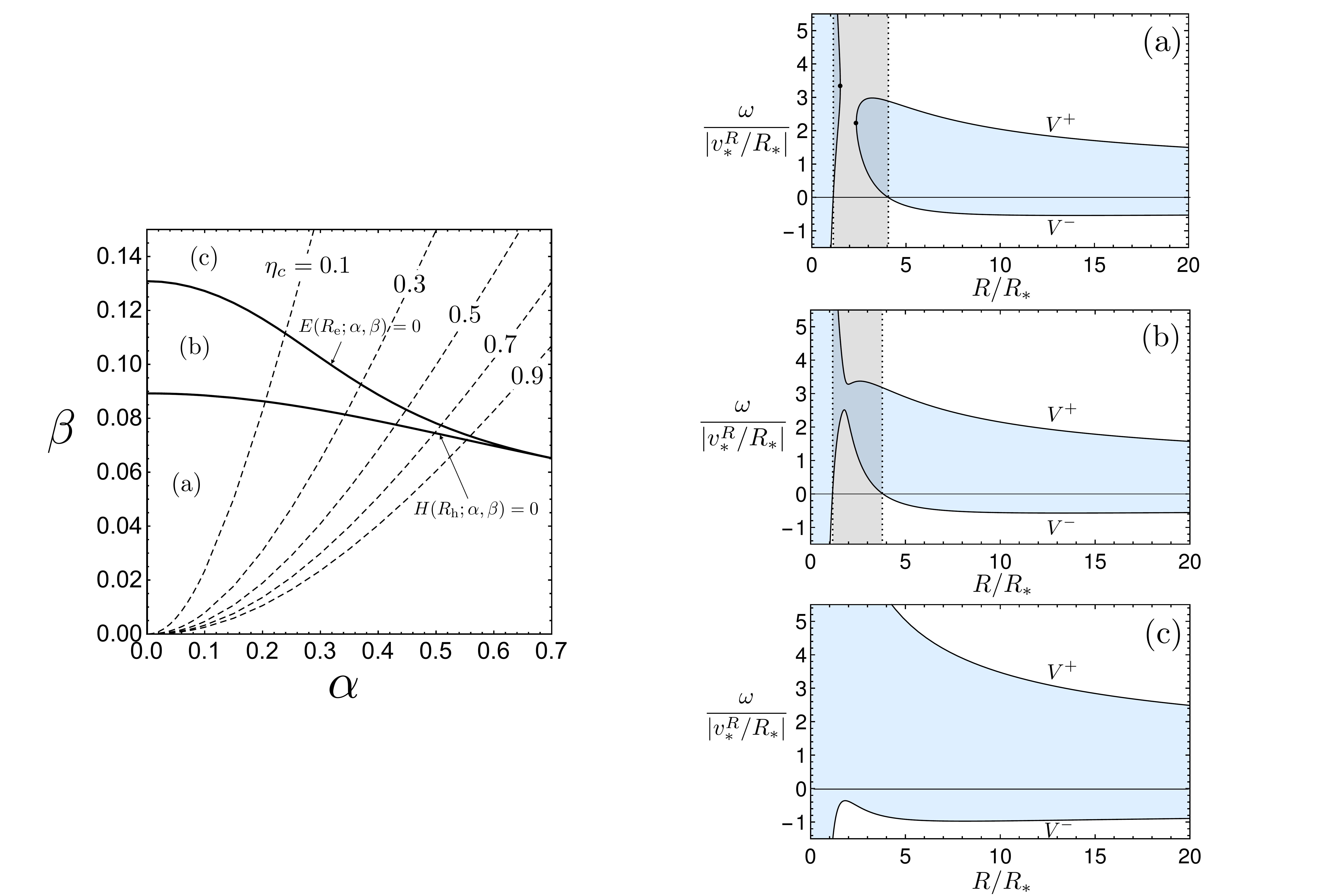}
\caption{The classification of the magnetoacoustic geometry for the
  fast mode with $\Gamma=4/3$ and $m=20$. In the left panel, the solid
  lines are $H(R_{\text{h}}(\alpha,\beta);\alpha,\beta)=0$
    (lower line) and $E(R_{\text{e}}(\alpha,\beta);\alpha,\beta)=0$
    (upper line).  The dashed lines are contours
  $\eta_c(\alpha,\beta)=0.1,\, 0.3,\, 0.5,\, 0.7,\, 0.9$.  There are
  three types---(a), (b), and (c)---of the effective potentials ${V}^{\pm}$
  classified by values of the parameters $(\alpha, \beta)$.  The right
  panels are typical effective potentials for
  (a):~$(\alpha,\beta)=(0.1,0.085)$; (b):~$(\alpha,\beta)=(0.1,0.089)$;
  and (c):~$(\alpha,\beta)=(0.1,0.14)$.  The grey regions in the right
  panels represent the magnetoacoustic ergoregions for the fast mode.}
\label{fig:fast_result}
\end{figure}

With the eikonal equation of the MHD waves, Eq.~\eqref{eq:eikonal_f}, we
introduce the magnetoacoustic rays as the integral curve of the vector
$k^\mu$ defined by
\begin{equation}
k^{\mu} \equiv\frac{dx^\mu}{d\lambda}= 
M^{\mu\nu}_{\text{fast}}\,\frac{\partial S}{\partial x^\nu},\quad
(M_{\text{fast}})_{\mu\nu}\,k^{\mu}k^{\nu}=0.
\label{eq:rays}
\end{equation}
%
In the magnetic pressure-dominated case, the magnetoacoustic metric 
\eqref{eq:acmetric_fast_tilde} becomes
\begin{equation}
\begin{aligned}
  ds_\text{fast}^2\propto
 &-\left[(V_\text{M}^2-
   \bm{v}^2)-\eta
 \, (\bm{b}\cdot  
 \bm{v})^2\right]d{t}^2
-2\left[ v^{\phi} +\eta\, b^{\phi}(\bm{b}\cdot
   \bm{v})
\right]
 R\, d{t}\,d{\phi}
\\
 &+
\frac{dR^2}{1
 -\eta\,(b^{R})^2-(v^R/V_\text{M})^2}
  +\left[1+\eta\,
   (b^{\phi})^2\right]R^2\,
   d{\phi}^2
  \label{eq:acmetric_fast_va_cs}
\end{aligned}
\end{equation}
with $\eta\approx\left(c_s/V_\text{A}\right)^2 \ll1$.
As the geometry defined by the metric
  \eqref{eq:acmetric_fast_va_cs} is stationary and axisymmetic, there
  exist two Killing vectors $\xi_{({t})}$ and
  $\xi_{({\phi})}$, and the conserved
  quantities associated with them are
\begin{equation}
\omega \equiv -k_{\mu}\,\xi^{\mu}_{(t)},\quad
 m \equiv k_{\mu}\,\xi^{\mu}_{(\phi)}.
\end{equation}
The phase of the eikonal equation is separated as 
$S=-\omega {t} + m{\phi}+S_{R}(R)$. 
We can obtain the equation for the radial motion of the magnetoacoustic 
rays in the same way  discussed  in Sec.~II for the acoustic rays. 
Then the eikonal equation \eqref{eq:rays} provides the following 
radial equation for magnetoacoustic rays:
\begin{equation}
\left(\frac{dR}{d\lambda}\right)^2=
\frac{1}{V_\text{A}^2}\left[1-\eta\,(b^R)^2\right]
(\omega-V^{+})(\omega-V^{-}),
\end{equation}
\begin{equation}
\begin{aligned}
    V^{\pm}&=m\frac{-(M_{\text{fast}})_{t \phi}\pm
    \sqrt{(M_{\text{fast}})_{t \phi}^2-(M_{\text{fast}})_{\phi \phi}
    (M_{\text{fast}})_{t t}}}{(M_{\text{fast}})_{\phi \phi}}\\
&=\frac{m}{R}\  \frac{v^{\phi}+\eta b^{\phi}(\bm{b}\cdot \bm{v}) \pm
    \sqrt{V_\text{M}^2-(v^R)^2
     -\eta\left[(v^R)^2-(b^{\phi})^2V_\text{M}^2\right]}}{1+\eta\,
   (b^{\phi})^2}.
  \label{eq:Vm}
\end{aligned}
\end{equation}
The magnetoacoustic rays are allowed to move in the $R$ region
determined by $\omega \leq V^{-}(R)$ or $V^{+}(R)\leq \omega$.  The zero
point of the square-root term in the effective potentials $V^{\pm}$
provides the condition for the magnetoacoustic horizon \eqref{eq:Mag_hor} as
\begin{align}
\nonumber 0&=V_\text{M}^2-(v^R)^2
     -\eta\left[(v^R)^2-(b^{\phi})^2V_\text{M}^2\right]\\
&\approx   V_\text{M}^2-(v^R)^2
     -\eta\,(b^R)^2V_\text{M}^2,
\end{align}
where we used the relation $V_\text{M}^2=(v^R)^2$ in the first-order
term of $\eta$.  Likewise, the condition for the
magnetoacoustic ergosurface \eqref{eq:Mag_ergo} is reproduced
from the condition $V^-=0$ by using  the relation $V_\text{M}^2=\bm{v}^2$ in the
first-order term of $\eta$.  As the components of the magnetoacoustic
metric $(M_{\text{fast}})_{\mu\nu}$ depend on $\alpha$ and $\beta$
through the background flow $\bm{v}$ and $\bm{B}$, the properties of
the effective potentials $V^{\pm}$ are determined by the parameter
sets $(\alpha, \beta)$. 
We classify the magnetoacoustic geometry by examining the existence of  
the magnetoacoustic horizon and ergoregion using conditions \eqref{eq:Mag_hor} 
and \eqref{eq:Mag_ergo}. Let us define the following functions of $R$: 
\begin{equation}
H(R;\alpha, \beta)\equiv 
   V_\text{M}^2-\eta(b^R)^2V_\text{M}^2-(v^R)^2\ \ ,\ \ 
E(R;\alpha, \beta)\equiv 
V_\text{M}^2-\eta (\bm{b} \cdot \bm{v})^2-\bm{v}^2.
\label{eq:H_E}
\end{equation} 
If these functions have intersections with the $R$ axis, the
magnetoacoustic geometry has the magnetoacoustic horizon and
ergoregion. The functions \eqref{eq:H_E} have one minimum at
$R_{\text{h}}=R_{\text{h}}(\alpha,\beta)$ and
$R_{\text{e}}=R_{\text{e}}(\alpha,\beta)$, which satisfy
$dH/dR|_{R_\text{h}}=0$ and $dE/dR|_{R_\text{e}}=0$.  Conditions for
functions \eqref{eq:H_E} to touch the $R$ axis are given by
\begin{equation} 
   H(R_{\text{h}}(\alpha,\beta);\alpha,\beta)=0\ \ ,\ \
 E(R_{\text{e}}(\alpha,\beta);\alpha,\beta)=0.
   \label{eq:boundary}\end{equation}
These relations \eqref{eq:boundary} determine boundaries
  between different types of geometries in the $\alpha\beta$ plane.
We show the results for $\Gamma=4/3$ in Fig.~\ref{fig:fast_result}
because we obtain the same classification of the magnetoacoustic
geometry for all $\Gamma$ within $0\leq \Gamma \leq 4$.
\noindent Figure \ref{fig:fast_result}\, shows the classification of possible
shapes of the effective potentials. For the validity of the
approximation adopted in our analysis, we focus only on the region
$\eta_c\le0.1$ in the left panel and we find the magnetoacoustic
geometries are classified into three types---(a), (b), and (c)---depending on the
existence of the magnetoacoustic horizon and the magnetoacoustic
ergoregion.  From the shape of the effective potentials $V^{\pm}$, we
see $R=0$ is surrounded by the potential wall with infinite height, and
the incident rays cannot reach the origin.  This behavior of the
effective potential is caused by the magnetic pressure; for small $R$, the 
$V_\text{M}^2$ term in the effective
potential \eqref{eq:Vm} becomes dominant and 
\begin{equation}
      V^{\pm}\approx\pm \frac{m}{R}\,V_\text{M}\approx\pm
      m\,\sqrt{1+\beta^2}~\frac{|v_*^R|}{R_*}\left(\frac{R}{R_*}\right)^{-7/4}.
  \end{equation}
  This explains the wall of the effective potential in the vicinity of
  $R=0$.  Although we only consider $\beta\ge0$ in
  Fig.~\ref{fig:fast_result}, it is possible to treat the $\beta<0$ region
  by changing the sign of the azimuthal quantum number $m$. Now, we
  discuss geometrical properties of each type.
  
  For type
  (a), both the magnetoacoustic horizon and the ergoregion exist. 
  Thus, the magnetoacoustic geometries are analogs of 
  rotating black holes. In this case, by checking the value 
  of the potentials at the magnetoacoustic horizon, we obtain the following 
  superradiant condition:
\begin{align}
\nonumber &\omega < 
\left.-m\frac{M^{\text{fast}}_{{t}{\phi}}}{M^{\text{fast}}_{{\phi}{\phi}}}
\right|_{R_\text{H}}
\approx m\frac{v^{\phi}}{R} 
\left. \left(1+\eta
            \left(\frac{v^R}{v^{\phi}}\right)
b^{R}\,b^{\phi}\right)\right|_{R=R_\text{H}}\\
&=-\frac{m\,\Omega_\text{F}}{R_\text{H}}\left[\left(\frac{R_\text{H}}{R_{*}}\right)^{{1}/{2}}+\left
  (\frac{R_\text{H}}{R_{*}}\right)^{-{1}/{2}}+1\right]\left. 
\left(1+\eta
  \left(\frac{v^R}{v^{\phi}}\right)
  b^{R}\,b^{\phi}\right)\right|_{R=R_\text{H}}~~\text{with}~
m\,\Omega_\text{F}<0.
\label{eq:superrad_MHD}
\end{align}
From the $t\phi$ component of the magnetoacoustic metric
\eqref{eq:acmetric_fast_va_cs}, the signs of $v^{\phi}$ and $b^{\phi}$
determine the direction of the black hole's spin.  As the signs of
$v^{\phi}$ and $b^{\phi}$ are the same and the sign of
$\Omega_{\text{F}}$ is opposite, for negative
$\Omega_{\text{F}}$ $(\beta>0)$, the magnetoacoustic black hole
rotates counterclockwise as shown in Fig.~\ref{fig:flow}.
Therefore, the additional condition $m\,\Omega_{\text{F}}<0$ means
that the incident magnetoacoustic rays should fall along prograde
orbits $m>0$ for superradiance.

 For the flow belonging to type (b), the magnetoacoustic horizon
disappears, and the potential wall becomes the inner
boundary of the magnetoacoustic ergoregion.  Hence, the flows are
analogs of the ultra-spinning compact stars which evoke the
ergoregion instability \cite{Friedman1978, Comins1978, Vilenkin1978,
  Cardoso2008, Glampedakis2013}.  The analog ergoregion instability
for the magnetoacoustic rays is possible, as rays are confined in the
magnetoacoustic ergoregion where rays have negative energies (the grey
region between two vertical dotted lines in the right panels of
Fig.~\ref{fig:fast_result}).  Finally, the magnetoacoustic rays can be
conveyed toward outside of the magnetoacoustic ergoregion with
positive energies via wave tunneling.  For type (c), superradiance
does not occur because there is no magnetoacoustic ergoregion.

\subsection{Slow mode}
As discussed in Sec. III, it is not possible to introduce the
magnetoacoustic metric for the slow mode. However, we can discuss the
motion of the magnetoacoustic rays from the eikonal equation
\eqref{eq:eikonal_s}. The directions of the propagation are given by
the characteristics
$\bm{v}_{\pm}= \bm{v}\pm (c_s V_\text{A}/V_\text{M}) \bm{b}
\approx \bm{v}\pm c_s \bm{b}$.
For the ingoing wave, $v^R_{-}=v^R-c_s\,b^R<0$ because $v^R<0$
and $b^R>0$ and the ingoing wave propagates towards $R=0$. On
the other hand, for the outgoing wave, $v_{+}^R$ is given by
\begin{equation}
  v_{+}^R=
  v^R+c_s b^R =|v^R_* |\left(\frac{R}{R_{*}}\right)^{-1/2}\left[-1+ \frac{\alpha }{\sqrt{1+\beta^2\left(1+\sqrt{R/R_*}\right)^2\left(1+R/R_*\right)^2}}\left(\frac{R}{R_{*}}\right)^{(3-\Gamma)/4}\right].
\label{eq:vp}
\end{equation}
For the polytropic index within $0\leq \Gamma \leq 3$, the limiting behavior of 
$v_{+}^R$ is
\begin{equation}
\lim_{R\rightarrow 0}v^R_{+}= -\infty, \quad
\lim_{R\rightarrow\infty}v^R_{+}=-0,
\label{eq:slow_vp}
\end{equation}
and we have checked that
$v^R_{+}<0$ for all $R$ under the condition $\eta_c<0.1$.
However, for $3 < \Gamma \leq 4$, the sign of $v^R_{+}$ 
becomes positive in the vicinity of the origin of the flow $R=0$, 
and hence there exists a point where $v^R_{+}=0$. 
Thus, the ``outgoing'' waves propagate towards this point from the outer region 
and wind around a circle because the azimuthal component $v^{\phi}_{+}$ is not zero 
even at this point. 
  Anyway, we do not have horizonlike structure
associated with the slow mode.

\section{concluding remarks}
We have discussed the magnetoacoustic geometry defined for the MHD
waves in an axisymmetric stationary inflow.  We found that for 
the magnetic pressure-dominated and the gas pressure-dominated
cases, the magnetoacoustic metric can be introduced only for the fast mode. 
For the slow mode, it is not possible to introduce effective
  geometries because its propagation is restricted along lines
determined by the background fluid flow and the magnetic field.  As
the background MHD flow, we considered the axisymmetric stationary
inflow characterized by two parameters $\alpha$ and $\beta$ specified
at the radial Alfv\'en point. Then, the property of the
magnetoacoustic geometry for this background flow can be classified
into three types corresponding to rotating black holes, ultraspinning
compact objects and rotating stars.

In order to have a complete understanding of the magnetoacoustic
geometry without the assumption adopted in this paper, we need to
introduce the magnetoacoustic metric for the general MHD wave equation.
Moreover, in our MHD inflow model, there is no Alfv\'en wave mode
since we have restricted our analysis to two-dimensional flows.
However, the Alfv\'en wave mode should be important when we consider 
three-dimensional MHD flows. In particular, in the high-energy astrophysical
phenomena (e.g., active galactic nuclei, gamma-ray bursts), we expect
that the study of the wave propagation in three-dimensional MHD flows
is essential to understand the energy transportation, the formation,
and evolution of some astrophysical systems. Therefore, analysis of
the magnetoacoustic geometries including the Alfv\'en mode in
astrophysical situations would provide us with important and interesting
topics.  We leave these problems to our future works.
\begin{acknowledgments}
  S.N. was supported by the Nagoya University Program of Leading Graduate 
  Schools funded by the Ministry of Education of Japanese Government with the 
  Program No. N01. 
  Y.N. was supported in part by JSPS KAKENHI Grant No. 15K05073. 
  M.T. was supported in part by JSPS KAKENHI Grant No. 24540268. 
\end{acknowledgments}



\end{document}